# THE DISPERSION RELATION FOR MATTER WAVES IN A TWO-PHASE VACUUM

Paul S. Wesson

[1]Department of Physics and Astronomy, University of Waterloo, Waterloo, ON,
 N2L 3G1, Canada.

Abstract:  The cosmological constant $\Lambda$ of general relativity is a natural consequence of embedding Einstein's theory in a five-dimensional theory of the type needed for unification.  The exact 5D solution for $\Lambda < 0$ shows waves in ordinary 3D space with properties similar to those of de Broglie or matter waves.  Here the dispersion relation is derived for matter waves in a toy two-phase model, where regions with $\Lambda < 0$ and $\Lambda > 0$ average on the large scale to $\Lambda \simeq 0$, thus providing in principle a resolution of the cosmological-constant problem.  A striking result of the analysis is that the dispersion relation is bimodal, with a well-defined window of high-frequency transmission which effectively defines the speed of light.

# THE DISPERSION RELATION FOR MATTER WAVES
# IN A TWO-PHASE VACUUM

1. <u>Introduction</u>

The classical vacuum is usually assumed to be uniform, but the quantum vacuum has energetic processes which imply inhomogeneity. That our understanding of the vacuum is incomplete is shown by the cosmological-constant problem. In Einstein's theory, the energy density of the vacuum is measured by the cosmological constant $\Lambda$, which can be positive or negative. However, the magnitude of this parameter is inferred to be large for particles and small for the observable universe, the mismatch being many orders of magnitude [1]. In principle, it is possible to resolve this problem with a model where different regions have energy densities for the vacuum as measured by $\Lambda$ with opposite signs, where the global average is close to zero. In such an inhomogeneous model, waves propagating through the vacuum are expected to show phase shifts or dispersion. In what follows, we will derive the dispersion relation for de Broglie or matter waves in a toy model with two phases.

Dispersion, where the phase speed depends on the wavelength or wave-number, can be shown by any kind of wave. It is well understood for electromagnetic waves [2]. It is less understood for de Broglie or matter waves, where dispersion causes the wave packet which accompanies a particle to spread [3-6]. However, it is really preferable to study these, because they show a richer variety of behaviour, and the photon case can always be recovered by letting the mass of the associated particle tend to zero. There has also been a recent upswing in the investigation of these waves in the laboratory [7-11].

Furthermore, for cosmological applications, an exact solution has recently been found which describes de Broglie waves propagating in a spacetime curved by a negative cosmological constant [12-14]. These waves have been termed "vacuum waves," but appear to be de Broglie or matter waves of the type which accompany particles moving through otherwise empty space.

The underlying solution is five-dimensional, and shows that the puzzling properties of de Broglie waves in four dimensions are due to truncating a well-behaved higher-dimensional metric. Solutions of this type are typical of Space-Time-Matter theory and Membrane theory, which are consistent with observations and may provide a route to the unification of gravity with the other fundamental interactions of particle physics [15-21]. The $\Lambda < 0$ solution of anti-de Sitter type for a wave is complemented by a well-known $\Lambda > 0$ solution of de Sitter type for a particle, so the combination might provide a model for wave-particle duality in agreement with the observation that spacetime on small scales is not severely curved. We wish to consider a similar 5D model for cosmology. This will not only allow us to study dispersion in the 5D vacuum, but also indicate how to improve on models for phase changes in the $\Lambda$ of 4D general relativity [22, 23] while helping to make contact with recent work on embeddings [24-26].

Our aim is to derive the dispersion relation for de Broglie waves and interpret it using the above-noted $\Lambda$-dominated solutions. We hope in this way to gain a better understanding of the possible properties of the vacuum. Everybody is familiar with the fact that the vacuum transmits light with its characteristic velocity, and indeed takes this for granted. We will find, though, that this may be the result of more complicated underlying behaviour. Our account is preliminary in nature, and after the working of Section 2 we will outline ways to develop it in Section 3.

2. Wave Dispersion and Properties of the Vacuum

We will use the same notation for the matter waves associated with particles of finite rest mass as for photons, and a brief review of nomenclature is in order. Thus $\omega$ is the frequency and $k$ is the wave number, but if the medium is inhomogeneous the phase velocity of the wave $v_p$ is not in general constant. It varies as $v_p = c/n(k)$ where $c$ is the velocity of light and $n(k)$ is the refractive index. (In terms of the permittivity or dielectric constant and the permeability, $n^2 = \varepsilon\mu$.) A wave packet or pulse of energy spreads as it propagates, due to its components moving at different speeds, and the velocity of energy flow is not the phase velocity but the group velocity $v_g$. For normal dispersion $dn/d\omega > 0$ and $v_g < v_p$. For anomalous dispersion $dn/d\omega < 0$ and $v_g > v_p$. For most ordinary substances, $n > 1$ and $v_p < c$. But even liquid water has a small region where $n(k) < 1$, reflecting its complicated microscopic structure. In general, complex values of $n$ (imaginary $\varepsilon$) indicate absorption, which however we will ignore in the present account.

For a particle with rest mass $m$, as opposed to a photon, de Broglie was led by a consideration of the relevant 4-vectors to infer that each component of the momentum $p$ has associated with it a wave with wavelength $h/p$, where h is Planck's constant. The mass itself may be represented by the Compton wavelength $h/mc$. (This can be regarded as the extra component of a 5-vector in 5D versions of relativity as discussed in ref. 7.)
The group velocity $v_g$ is identified with the ordinary velocity of the particle [6]. An integral part of de Broglie's formulation of wave mechanics is that the product of $v_p$ and $v_g$ for matter waves is equal to $c^2$. This is consistent with a certain interpretation of the Lorentz transformations [27], but naturally raises questions about causality (see below). However, modern versions of the double-slit experiment, as well as extensive investigations with neutron interferometry, leave little doubt about the validity of de Broglie's approach [7-11]. We therefore proceed on the assumption that de Broglie waves provide an accurate description of things, at least on the small or local scale.

The considerations of the preceding paragraph can be summed up in a few equations:

$$\omega = ck/n, \quad \omega = \omega(k), \quad n = n(k) \tag{1}$$

$$v_p = \omega/k = c/n \tag{2}$$

$$v_g = \frac{d\omega}{dk} = \frac{c}{n + \omega(dn/d\omega)}. \tag{3}$$

In the last of these, the two forms are equivalent, as may be verified by taking the derivative with respect to $k$ of (1) in the form $\omega n = ck$. To the above must be added the de Broglie relation noted above:

$$v_p v_g = c^2. \tag{4}$$

It is worthwhile noting that while (1)-(4) are familiar, they are the result of the evolution of our knowledge of wave phenomena through history, and do not form a logically complete set of relations. For example, it has been remarked before that there are ambiguities in the dispersion relation $\omega = \omega(k)$ for de Broglie waves [6]. This can be appreciated by noting that in the four equations above there appear six unknowns, so it is impossible to uniquely specify any of them without further information, such as might be provided in a practical situation by boundary conditions. In cosmology, of course, we do not have access to boundary conditions, so we perforce proceed in a theoretical mode using (1)-(4) as a basis. It should be noted that (1)-(3) and (4) are normally treated separately, so in combining them as we do below, we may expect some new behaviour to appear.

The dispersion of vacuum waves can be studied by obtaining a relation for the effective refractive index as a function of frequency. To do this, we substitute from (2) and the second member of (3) into (4). There results

$$\frac{dn}{d\omega} = \frac{1-n^2}{n\omega} \qquad (5)$$

This shows that what are commonly called normal dispersion ($dn/d\omega > 0$) and anomalous dispersion ($dn/d\omega < 0$) are separated by $n=1$ for de Broglie waves, just as for electromagnetic waves. For the two ranges we find by integrating (5) that

$$n = \left[1 - \left(\frac{\omega_0}{\omega}\right)^2\right]^{1/2}, \quad n < 1 \qquad (6.1)$$

$$n = \left[1 + \left(\frac{\omega_0}{\omega}\right)^2\right]^{1/2}, \quad n > 1 \qquad (6.2)$$

Here $\omega_0$ is an arbitrary constant that fixes the dispersion relations derived below for both ranges of $n$. For $n < 1$, $\omega_0$ is akin to the cutoff frequency in a waveguide, where the physical size of the device means that waves with lengths greater than a critical value cannot propagate, or that frequencies below a critical value cannot propagate. Likewise for de Broglie waves, where for $n<1$ only frequencies $\omega > \omega_0$ can propagate. This means that for the vacuum there is no intrinsic high-$\omega$ cutoff, so the magnitude of the energy density ($\sim k^4$ or $\omega^4$) can be large [28]. Indeed, if the practical cutoff is at the Planck scale, the energy density is of the order $10^{112}$ erg cm$^{-3}$. By contrast, the energy density corresponding to the observed cosmological constant is of order $10^{-8}$ erg cm$^{-3}$. This is the cosmological-constant problem mentioned in Section 1.

To obtain explicit dispersion relations for $n<1$ and $n>1$, we take (6.1) and (6.2) in the form $1-n^2 = \pm(\omega_0/\omega)^2$, and replace $n$ by $ck/\omega$ using (1). Multiplying throughout by $\hbar^2$ (where $\hbar \equiv h/2\pi$) gives

$$(\hbar\omega)^2 = (c\hbar k)^2 \pm (\hbar\omega_0)^2 \quad , \qquad (7)$$

where on the right-hand side the upper sign refers to $n<1$ and the lower sign to $n>1$. These are dispersion relations for vacuum or de Broglie waves.

We can also express (7) in terms of mechanical quantities by making the appropriate transitions to energy, 3-momentum and rest mass ($E = \hbar\omega$, $p = \hbar k$, $m_0 = \hbar\omega_0/c^2$). Then (7) reads

$$E^2 = p^2 c^2 \pm m_0^2 c^4 \qquad . \qquad (8)$$

This relation with the upper sign ($n<1$) is the standard normalization condition of quantum mechanics for the energy, momentum and mass when the last is "on shell". It is also the relation found in special or general relativity when the proper time is given in terms of the metric tensor by $ds^2 = g_{\alpha\beta} dx^\alpha dx^\beta$ and the 4-velocities $u^\alpha \equiv dx^\alpha/ds$ are normalized via $u^\alpha u_\alpha = 1$, which when expanded and multiplied throughout by a real constant $m_0^2$ gives (8). By contrast, the lower sign in (8) is associated with masses which are "off shell", a possibility raised by the approach to quantum mechanics due to Stueckelberg and Feynman [29-34]. In the language of special relativity, particles obeying (8) with the lower sign would be called tachyons. However, this is somewhat simplistic. In fact, (8) with the lower sign is equivalent to the null path of a particle in a 5D manifold with a timelike extra dimension [15, 35-37]. Such manifolds are used in Space-Time-Matter theory and Membrane theory.

This is not the place to discuss the details of these theories, but a few comments are in order. The noted theories were devised in order to explain the origin of matter and the masses of particles, which general relativity does not. However, the 5D theory contains 4D general relativity, by virtue of an embedding theorem due to Campbell. The field equations are commonly taken to be given in terms of the 5D Ricci tensor by

$$R_{AB} = 0 \quad (A, B = 0,123,4) \quad . \quad (9)$$

There are physically acceptable solutions of these equations for both a spacelike and timelike extra dimension. When the 5D metric takes the so-called canonical form, *all* 4D solutions of Einstein's theory with a cosmological constant $\Lambda$ are embedded [20, 21]. When the extra coordinate $x^4 = l$ is spacelike, it is found that $\Lambda > 0$ and the null geodesics of the 5D metric are monotonic or particle-like. When $x^4 = l$ is timelike, $\Lambda < 0$ and the null geodesics are oscillatory or wave-like. In general, 5D geodesics which are null correspond to the 4D paths of *both* massless photons and massive particles, so $dS^2 = 0$ defines higher-dimensional causality.

The exact solution of (9) which gives the 5D analog of the 4D cosmological de Sitter solution with $\Lambda > 0$ has been known for many years [15]. It has been discussed at length in the literature, so we only quote it:

$$dS^2 = \frac{l^2}{L^2}\left\{c^2 dt^2 - \exp\left[\frac{2ct}{L}\right](dx^2 + dy^2 + dz^2)\right\} - dl^2 \quad . \quad (10)$$

Here the constant length $L$ measures the scale of the 4D potential, and by the reduction of (9) to Einstein's equations can be shown to be related to the cosmological constant by $\Lambda = 3/L^2$.

The exact solution of (9) with $\Lambda < 0$ is of more recent discovery [12]. However, solutions of similar type have appeared occasionally in the literature, and are typified by having complex metric coefficients [15]. This is admissible, provided that the physical quantities calculated from the field equations are real. This applies especially in the present case to the value of the cosmological constant. The easiest way to verify this, and to check the solution, is by computer [1]. Then it may be shown that an exact solution of (9) is given by:

---

[1] The computer may be employed not only to verify the solution (11) of the main text, but also to show that it only exists in the noted form if the extra dimension is timelike. The latter causes $\Lambda < 0$, so as noted before the density and pressure of the vacuum are negative and positive, respectively. These are the opposite of ordinary matter, which causes some differences in optical properties (compare ref. 2 and Fig. 1). For ordinary matter, normal dispersion is defined by $dn/d\omega > 0$ and $n > 1$, while anomalous dispersion has $dn/d\omega < 0$ and $n < 1$. For timelike vacuum, these dependencies are reversed.

$$dS^2 = \frac{l^2}{L^2}\left\{c^2dt^2 - \exp\left[\pm\frac{2i}{L}(ct+\alpha x)\right]dx^2 - \exp\left[\pm\frac{2i}{L}(ct+\beta y)\right]dy^2\right.$$

$$\left. - \exp\left[\pm\frac{2i}{L}(ct+\gamma z)\right]dz^2\right\} + dl^2 \quad . \tag{11}$$

In this case, as opposed to (10), it can be shown that $\Lambda = -3/L^2$. There is a wave in 3D, whose frequency is $f = 1/L$ and whose wave-numbers in the three directions of ordinary space are $k_x = \alpha/L$, $k_y = \beta/L$, $k_z = \gamma/L$. The dimensionless constants $\alpha, \beta, \gamma$ are arbitrary, so the speed of the wave in the x-direction (say) is $c/\alpha$ and can exceed $c$. Other properties of the wave show that it is of de Broglie type, and obeys his relation (4). The wave exists because of the vacuum (it is not a conventional gravitational wave). Since the equation of state of the classical vacuum is $p_v = -\rho_v c^2 = -\Lambda c^4/8\pi G$, we see that the wave described by (11) propagates in a fluid with negative density and positive pressure.

The solution (11) gives insight to the dispersion relation (7). Indeed, we infer that de Broglie waves may propagate in the vacuum. Then the two modes of dispersion shown by (7) suggest that the vacuum may have two phases, with $\Lambda > 0$ and $\Lambda < 0$. A two-phase model for the vacuum will require detailed work which goes beyond the scope of the present account. However, several comments can be made now which are based on the averaging of a variable vacuum energy density [1] and the application of 5D solutions to 4D wave-particle duality [12, 35-37]. The latter phenomenon is a fundamental property of matter, and can naturally be accommodated by combining two canonical solutions like (10) and (11) above. This can be achieved algebraically by generalizing the canoni-cal metric so as to include a complex scalar field $g_{44}(x^\gamma, l)$ whose two modes have the effect of switching the signature [15]. At small scales, this leads to an exact balance, since every particle has an associated wave with the opposite sign of $\Lambda$. Put another way, with reference to the normalization relation (8): each particle has a kind of 'charge' dependent on the square of its inertial mass $+m_0^2$, whereas the associated wave has an effective charge $-m_0^2$. Of course, this matching on the microscopic scale cannot be expected to hold on the macroscopic scale. At some large scale the influence of the particles becomes lost, and the gravitational scale of $10^{28}$ cm becomes dominant, resulting in a global value of $\Lambda \sim 10^{-56}$ cm$^{-2}$ as observed.

To return to the main theme, we note that the bimodal nature of the dispersion relation (7) comes from the bifurcation in the refractive index at $n=1$ as shown in (6.1) and (6.2). We have plotted the refractive index and frequency in Fig. 1. (The refractive index in the present context is phenomenological and largely the result of history.) We have also plotted the phase and group velocities, according to (2) and (4). When plots are made in this manner, it is clear that the medium has some kind of structure which makes itself evident at $n=1$, when the phase and group velocities both equal the speed of light. There is a divergence in $\omega$ at $n=1$, indicating that arbitrarily high frequencies can propagate there, or that arbitrarily large amounts of energy can pass.

3. <u>Conclusion</u>

The transmission of energy by an inhomogeneous medium is governed by the dispersion relation, which for the vacuum is given by (7) and is striking in being bimodal. This reflects the choice of $\Lambda > 0$ or $\Lambda < 0$ in two complementary solutions of the field equations which embed the two versions of 4D de Sitter space in 5D, where we have concentrated on the $\Lambda < 0$ solution (11) because the focus is on de Broglie waves [12, 35-37]. The two choices of $\Lambda$ are relevant to a model universe with regions of energy density with opposite sign, which might average to a finite but small value and thereby help resolve the cosmological-'constant' problem. The physical regions with differing signs of $\Lambda$ correspond to parts of the dispersion plot separated by an effective refractive index of unity, as illustrated in Fig 1. The sharp peak in the refractive index at unity corresponds to where arbitrarily high frequencies can be transmitted, the velocity there being that of light. This behaviour reproduces the well-known fact that the velocity of light is unique.

Our model is preliminary, and only outlines what is expected if the vacuum is indeed inhomogeneous. A more sophisticated model would specify the spectrum of the inhomogeneities, both in terms of physical size and the magnitude of the energy densities concerned. Also, relevant new information has become available recently on de Broglie or matter waves from redoing the double-slit experiment with modern technology and from neutron interferometry with improved equipment [7-11]. It seems to us that the dispersion of de Broglie waves is a promising way to probe the vacuum on small scales, and so gain information that may be relevant to the vacuum on large scales.

Acknowledgements

This project comes out of work done with other members of the Space-Time-Matter group (5Dstm.org), especially J.M. Overduin who checked the solution (11) on the computer. Thanks for comments on neutron interferometry and de Broglie waves go to S. Werner.

References

[1]  T. Padmanabhan, Phys. Rep. 380 (2003) 235.

[2]  J.D. Jackson, Classical Electrodynamics, 2nd. ed. (1975) 285, 302.

[3]  L. de Broglie, Ph.D. Thesis, Un. Sorbonne, Paris (1924).

[4]  L. de Broglie, Philosoph. Mag. 47 (1924) 446.

[5]  L. de Broglie, Ann. Phys. (Paris) 2 (1925) 10.

[6]  D. Paul, Am. J. Phys. 48 (1980) 283.

[7]  S. Kocsis, B. Braverman, S. Roberts, M.J. Stevens, R.P. Mirin, L.K. Shalm, A.M. Steinberg, Science 332 (2011) 1170.

[8]  K. Edamatsu, R. Shimizu, T. Itoh, Phys. Rev. Lett. 89 (2002) 213601.

[9]  D.M. Greenberger, W.P. Schleich, E.M. Rasel, Phys. Rev. A. 86 (2012) 063622.

[10] R. Colella, A.W. Overhauser, S.A. Werner, Phys. Rev. Lett. 34 (1975) 1472.

[11]  H. Rauch, S.A. Werner, Neutron Interferometry, Clarendon Press, Oxford (2000).

[12]  P.S. Wesson, Phys. Lett. B 722 (2013) 1, arXiv: 1301.0333 (2013).

[13]  P.S. Wesson, J.M. Overduin, arXiv: 1302.1190 (2013).

[14]  P.S. Wesson, arXiv: 1205.4452 (2012).

[15]  P.S. Wesson, Five-Dimensional Physics, World Scientific, Singapore (2006).

[16]  P.S. Wesson, J. Ponce de Leon, J. Math. Phys. 33 (1992) 3883.

[17]  P.S. Wesson, Gen. Rel. Grav. 40 (2008) 1353.

[18]  L. Randall, R. Sundrum, Mod. Phys. Lett. A 13 (1998) 2807.

[19]  N Arkani-Hamed, S. Dimopoulos, G. Dvali, Phys. Lett. B 429 (1998) 263.

[20]  B. Mashhoon, H. Liu, P.S. Wesson, Phys. Lett. B 331 (1994) 305.

[21]  S.S. Seahra, P.S. Wesson, Class. Quant. Grav. 20 (2003) 1321.

[22]  R.N. Henriksen, A.G. Emslie, P.S. Wesson, Phys. Rev. D 27(1983) 1219.

[23]  P.S. Wesson, Phys. Rev. D 34(1986) 3925.

[24]  C.M. Ho et al., Mod. Phys. Lett. A 28 (2013) 1350005.

[25]  T.J. Gao, T-F. Feng, J.B. Chen, Mod. Phys. Lett. A 27 (2012) 1250011.

[26]  P. Moravek, J. Hoeji, Mod. Phys. Lett. A 27 (2012) 1250098.

[27] W. Rindler, Essential Relativity, 2nd. ed., Springer, New York (1977) 91.

[28] S. Carroll, Spacetime and Geometry, Addison-Wesley (2004) 173.

[29] E.C.G. Stueckelberg, Helv. Phys. Acta. 14 (1941) 588.

[30] E.C.G. Stueckelberg, Helv. Phys. Acta. 15 (1942) 23.

[31] R.P. Feynman, Phys. Rev. 80 (1950) 440.

[32] I. Aharanovich, L.P. Horwitz, J. Math. Phys. 47 (2006) 1229023.

[33] I. Aharanovich, L.P. Horwitz, J. Math. Phys 51 (2010) 052903.

[34] I. Aharanovich, L.P. Horwitz, J. Math. Phys 52 (2011) 082901.

[35] P.S. Wesson, Phys. Lett. B 538 (2002) 159.

[36] P.S. Wesson, Phys. Lett. B 701 (2011) 379.

[37] P.S. Wesson, Phys. Lett. B 706 (2011) 1.

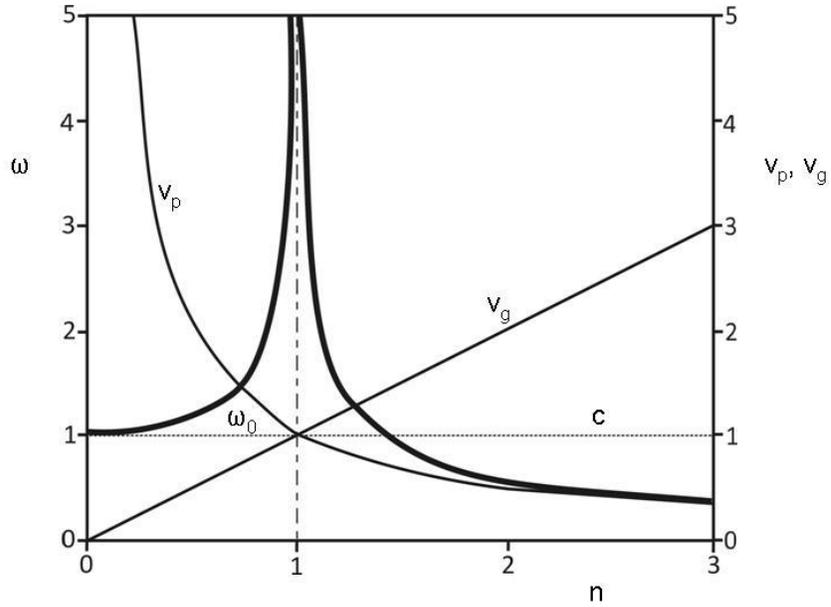

Fig. 1. Plots of relevant quantities as a function of the effective refractive index *n* for de Broglie waves. The regular lines show the phase velocity $v_p$ and group velocity $v_g$ according to equations (2) and (4), where the speed of light is *c*. The heavy lines show the relationship between the refractive index and the frequency $\omega$ according to equations (6.1) and (6.2), where $\omega_0$ is the cutoff frequency.